\documentclass[runningheads]{llncs}

\usepackage{graphicx}
\usepackage{tabularx}
\usepackage{cite}

\usepackage{academicons}
\usepackage{hyperref}
\usepackage{xcolor}
\newcommand{\orcid}[1]{\href{https://orcid.org/#1}{\includegraphics[scale=0.08]{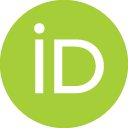}}}

\begin{document}
\title{The S$^+$($^4$S) + SiH$_{2}$($^1$A$_1$) reaction: toward the synthesis of interstellar SiS }
\titlerunning{The S$^+$($^4$S) + SiH$_{2}$($^1$A$_1$) reaction}
%
\author{Luca Mancini\inst{1}\orcid{0000-0002-9754-6071}
\and  Marco Trinari\inst{1}
\and Emília Valença Ferreira de Aragão\inst{1,2}\orcid{0000-0002-8067-0914} 
\and Marzio Rosi\inst{3}\orcid{0000-0002-1264-3877} 
\and  Nadia Balucani\inst{1}\orcid{0000-0001-5121-5683}
}

\authorrunning{L. Mancini et al.}
%
\institute{
Dipartimento di Chimica, Biologia e Biotecnologie,\\ Universit\`{a} degli Studi di Perugia, 06123 Perugia, Italy\\ 
\email{\{emilia.dearagao,luca.mancini2\}@studenti.unipg.it}\\
\email{nadia.balucani@unipg.it}\\
\and
Master-tec srl, Via Sicilia 41, 06128 Perugia, Italy\\
\email{emilia.dearagao@master-tec.it}\\
\and
 Dipartimento di Ingegneria Civile ed Ambientale,\\ Universit\`{a} degli Studi di Perugia, 06125 Perugia, Italy\\
 \email{marzio.rosi@unipg.it}
}

\maketitle              
\begin{abstract}
We have performed a theoretical investigation of the S$^+$($^4$S) + SiH$_{2}$($^1$A$_1$) reaction, a possible formation route of the HSiS$^+$ and SiSH$^+$ cations that are alleged to be precursors of interstellar silicon sulfide, SiS. 
Electronic structure calculations allowed us to identify the main reaction pathways for the systems. The reaction has two exothermic channels leading to the isomeric species $^3$HSiS$^{+}$ and $^3$SiSH$^{+}$ formed in conjunction with H atoms. The reaction is not characterized by an entrance barrier and, therefore, it is believed to be fast also under the very low temperature conditions of interstellar clouds. The two ions are formed in their first electronically excited state because of the spin multiplicity of the overall potential energy surface.
In addition, following the suggestion that neutral species are formed by proton transfer of protonated cations to ammonia, we have derived the potential energy surface for the reactions $^3$HSiS$^{+}$/$^3$SiSH$^{+}$ + NH$_{3}$($^{1}$A$_1$). 
\end{abstract}
\section{Introduction}
Silicon, the seventh most abundant element in the universe, is mostly trapped in the dust grains present in the interstellar medium (ISM) in the form of silicates. Nevertheless, an increasing number of silicon-bearing molecules, have been detected in different regions of the ISM, with a  molecular size going up to 8 atoms ~\cite{cdms}. The presence of silicon in the gas phase is mainly related to violent events, such as shocks, in which the high energetic material ejected from young stars impact a quiescent zone and allows the sputtering of the refractory core of interstellar dust grains~\cite{podio2017silicon}. Immediately after the release of  silicon in the gas phase, it can immediately react and it is mostly converted into SiO ~\cite{herbst1989chemistry,mackay1995chemistry}. Therefore, the diatomic molecule SiO is the most abundant silicon-bearing species and it is considered a useful target to probe shock regions ~\cite{podio2017silicon}. Formation routes of the SiO molecule are well characterized ~\cite{schilke1997sio,le2001si,gusdorf2008sio}, while chemical processes leading to the formation of other Si-bearing molecules are still uncertain and poorly known.\\
\indent
Another interesting Si-bearing diatomic molecule is silicon sulfide, SiS, detected for the first time in 1975 by Morris et al. ~\cite{morris1975detection} towards the molecular envelope of
the carbon-rich AGB star IRC+ 10216. From that first observation, silicon sulfide (SiS) has been detected
sporadically, mostly in circumstellar envelopes around evolved stars \cite{cernicharo2000lambda2,prieto2015si} and towards high mass star-forming
regions with shocks (such as Sgr B2 and Orion KL \cite{dickinson1981interstellar,ziurys1988sis,ziurys1991sis,tercero2011line,mackay1995chemistry}).
Recently, a census of silicon molecules has been performed by Podio et al. for L1157-B1, that is,  
a shocked region driven by a jet originated from a Sun-like protostar \cite{podio2017silicon}. Together with the detection of SiO (and its isotopologues $^{29}$SiO and $^{30}$SiO), the authors reported the first detection of the SiS molecule in a low-mass star forming region. 
Interestingly, an unexpected high
abundance of SiS has been inferred in a well-localized region around the protostar, with a SiO/SiS abundance ratios decreasing from typical values of 200 to values as low as 25. Since the abundance of other sulfur-bearing species (e.g. SO) remains constant across the region, the reason  for such diversity cannot be ascribed to a local overabundance of sulphur. The analysis of the spatial distribution shows a strong gradient across the shock (being SiS abundant at the head of the cavity and not detected at the shock impact region), suggesting a different chemical origin for SiO and SiS.
The reason for such a peculiar diversity is unknown and represents a challenge to our comprehension of interstellar silicon chemistry.\\
\indent
Given the difficulties associated to the experimental investigation of the postulated reactions, theoretical quantum chemistry calculation appears to be pivotal to elucidate the chemistry of silicon-bearing species. 
Recently, some interesting analysis on the potential role of neutral gas-phase chemistry in the formation
and destruction routes of SiS has been proposed. For instance, some of the
present authors performed a theoretical characterization of the potential energy surface (PES) of the reactions SiH+S, SiH+S$_2$ and Si+HS \cite{rosi2018possible,rosi2019electronic}, together with a kinetic analysis based on  the capture theory and Rice-Ramsperger-Kassel-Marcus calculations\cite{skouteris2018theoretical}.
In addition, Zanchet et al. \cite{zanchet2018formation} examined the reaction of atomic silicon with SO and SO$_2$ and used the derived rate coefficients to simulate SiS formation in the outflows of star-forming regions like Sgr B2, Orion KL, and L1157-B1. In particular, the PES of the SiOS system has been analyzed through ab initio calculations. Kinetics calculations on the derived PES lead to the conclusion that the two reactions can play a key role for the formation of SiS, while the reaction of silicon sulfide with atomic oxygen represents an important destruction pathway of SiS. 
Very recently, after the work by Rosi et al. \cite{rosi2019electronic}, the reaction of the silicon atom with the SH radical has been re-investigated by Mota et al. \cite{mota2021sis}. Accurate multi-reference calculations were employed to derive the potential energy surface for the system and quasiclassical trajectory calculations were performed to derive the reaction rate coefficient. 
The inclusion of the data in a gas-grain astrochemical model of the L1157-B1 shock region, suggested a very important role for the SiS formation.\\
\indent
One last example revealing the pivotal role of neutral gas phase chemistry for the formation of interstellar silicon sulfide is represented by the work of  R. Kaiser and coworkers, reporting a synergistic theoretical and experimental analysis of different reaction involving silicon
species\cite{doddipatla2021nonadiabatic,goettl2021crossed}. The combined crossed molecular beam and electronic
structure calculations analysis provided new data on the formation of silicon sulfide. The synergy with astrochemical modeling allowed a better understanding of the role of neutral gas-phase chemistry toward star
forming regions.\\
\indent
In spite of the recent advances summarized above, the two main databases for astrochemical models (KIDA and UMIST)\cite{wakelam20152014,mcelroy2013umist} still report the electron-ion recombination of the HSiS$^+$ ion as the only significant process leading to the formation of SiS:
\begin{equation}
      HSiS^{+} + e^{-} \rightarrow H + SiS   \\
   \end{equation}
However, in those databases most of the ion-molecule reactions considered as possible formation routes of the HSiS$^{+}$ cation are inefficient according to laboratory experiments\cite{wlodek1989gas,wlodek1987gas}. \\
\indent 
As a part of our systematic investigation on interstellar SiS formation, we have now focused on possible ion-molecules reactions. In particular, here we report on the theoretical characterization of the PES for the reaction between the S$^+$ cation (in its ground state, $^4$S) and the SiH$_{2}$($^1$A$_1$) radical. Sulphur cations are easily formed because the sulphur ionization energy (10.36 eV) is lower than that of atomic or molecular hydrogen. The SiH$_{2}$ radicals can be formed by photodissociation or other high-energy processes involving SiH$_4$. This last species can be synthesized by successive hydrogenation processes of Si atoms on the surface of the interstellar dust grains \cite{mackay1995chemistry,ceccarelli2018evolution} and subsequently released in the gas phase during shocks. \\
\indent
In addition to electron-ion recombination, other processes can be invoked for the formation of neutral molecules starting from a protonated charged species, that is, proton transfer to neutral molecules with a large proton affinity like ammonia. As proposed by Taquet et al. \cite{taquet2016formation} proton transfer to ammonia can be considered a good alternative to electron-ion recombination.
Therefore, we have also derived the PES for 
\begin{equation}
      HSiS^{+}/SiSH^{+} + NH_3 \rightarrow SiS + NH_4^+   \\
   \end{equation}

With the present contribution we aim to add other new pieces to the complex puzzle of the chemistry of interstellar silicon.

\section{Computational details}
The SiH$_{2}$($^1$A$_1$) + S$^+$($^4$S) and HSiS$^{+}$/SiSH$^{+}$ + NH$_{3}$($^{1}$A$_1$) reactions were analyzed adopting a computational strategy previously used with success in several cases \cite{rosi2018possible,skouteris2019interstellar,mancini2021reaction,balucani2018theoretical}. A first investigation of the potential energy surface was performed through Density Functional Theory (DFT) calculations, using B3LYP \cite{becke1993new,stephens1994ab} hybrid functional, in conjunction with the correlation consistent valence polarized set aug-cc-pV(T+d)Z \cite{dunning1989gaussian,woon1993gaussian,kendall1992electron}. Harmonic vibrational frequencies were calculated at the same B3LYP/aug-cc-pV(T+d)Z level of theory, using the Hessian
matrix (second derivatives) of the energy, in order to determine the nature of each stationary
point, i.e. minimum if all the frequencies are real and saddle point if there is one, and only one, imaginary frequency. Intrinsic Reaction Coordinates (IRC) calculations\cite{gonzalez1989improved,gonzalez1990reaction} were performed to assign each identified stationary point to the corresponding reactants and products. Subsequently, single points energy calculations, using the CCSD(T)/aug-cc-pV(T+d)Z level of theory \cite{bartlett1981many,raghavachari1989fifth,olsen1996full}, were performed on the optimized geometry of each stationary point. The zero-point energy, computed using the scaled
harmonic vibrational frequencies obtained at the B3LYP/aug-ccpV(T+d)Z level of theory, was added to correct both energies [B3LYP and CCSD(T)] to 0 K. All calculations were carried out using GAUSSIAN 09
\cite{frisch2009gaussian} and the vibrational analysis was performed using MOLDEN \cite{schaftenaar2000molden,schaftenaar2017molden}

\section{Results}
\subsection{The SiH$_{2}$($^1$A$_1$) + S$^+$($^4$S) reaction}

\begin{figure}[h!]
\includegraphics[width=1\linewidth]{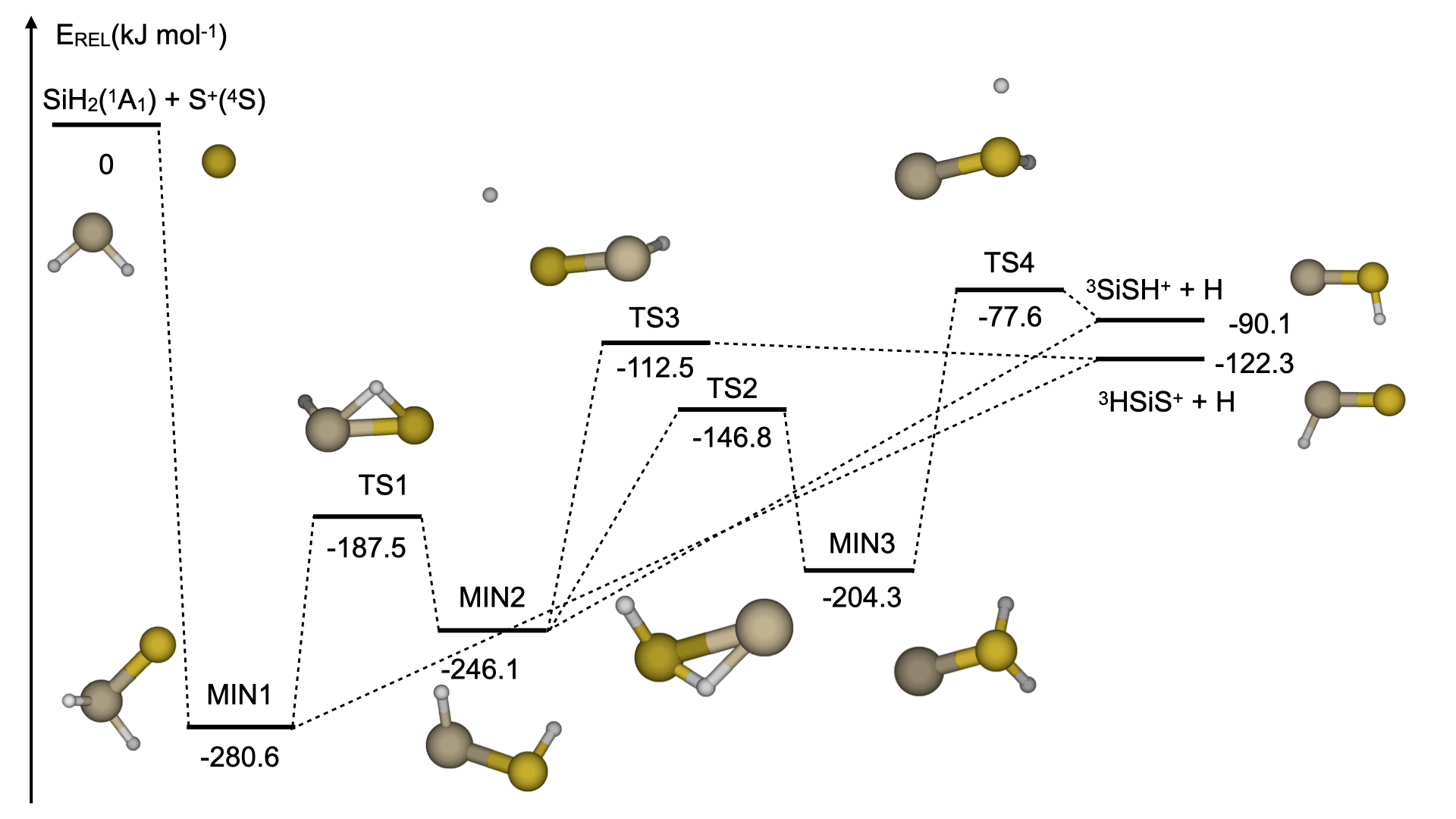}
\caption{2D scheme of the Potential Energy Surface (PES) for the reaction SiH$_{2}$($^1$A$_1$) + S$^+$($^4$S) with the energies evaluated at the CCSD(T)/aug-cc-pV(T+d)Z level of theory. } \label{fig1}
\end{figure}

\begin{figure}[h!]
\centering
\includegraphics[width=0.5\linewidth]{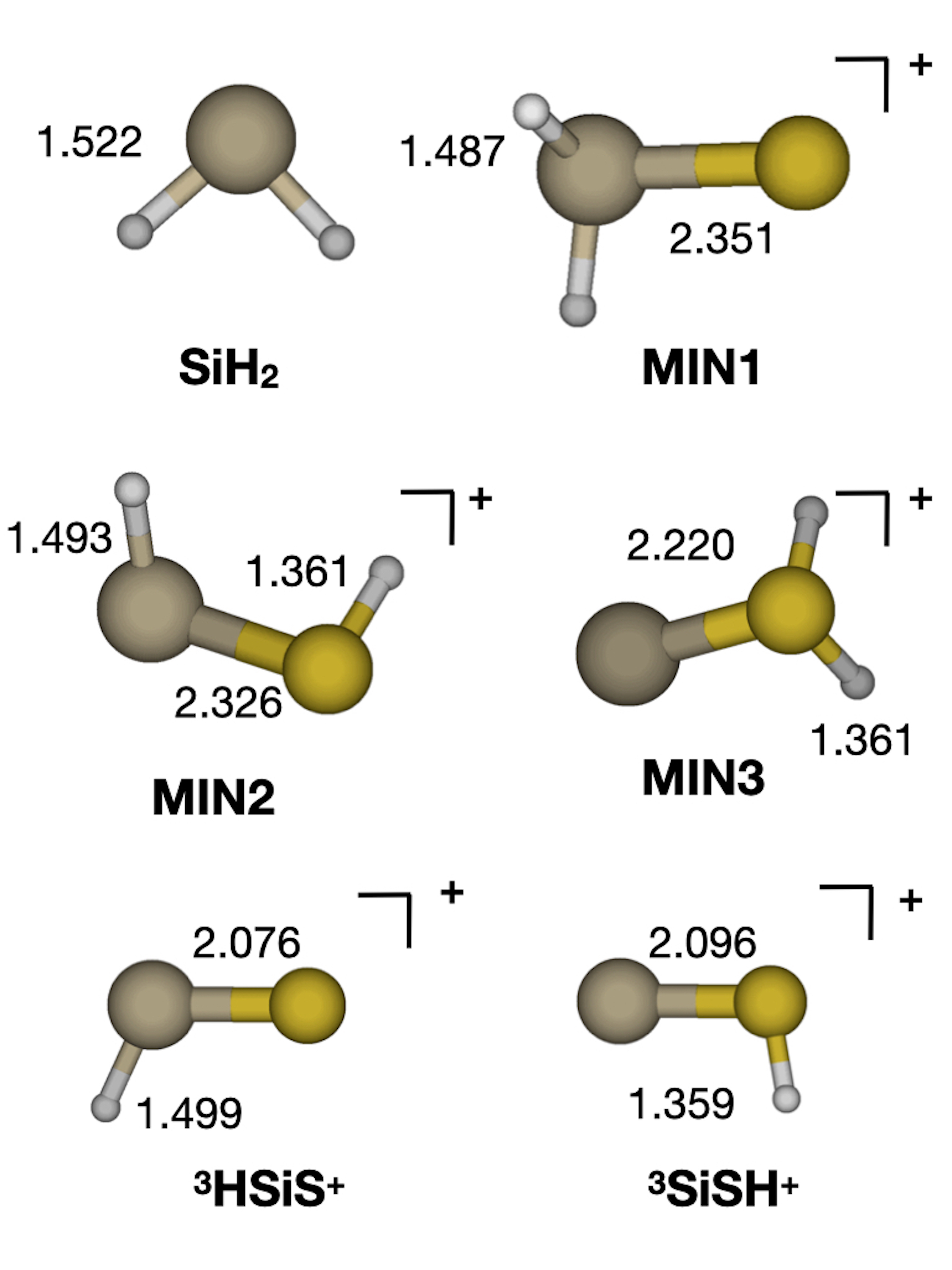}
\caption{Bond distances (reported in Angstroms) for the reactant, products and minima identified for the  SiH$_{2}$($^1$A$_1$) + S$^+$($^4$S) reaction at the B3LYP/aug-cc-pV(T+d)Z level of theory.  } \label{fig2}
\end{figure}

\begin{figure}[h!]
\centering
\includegraphics[width=0.5\linewidth]{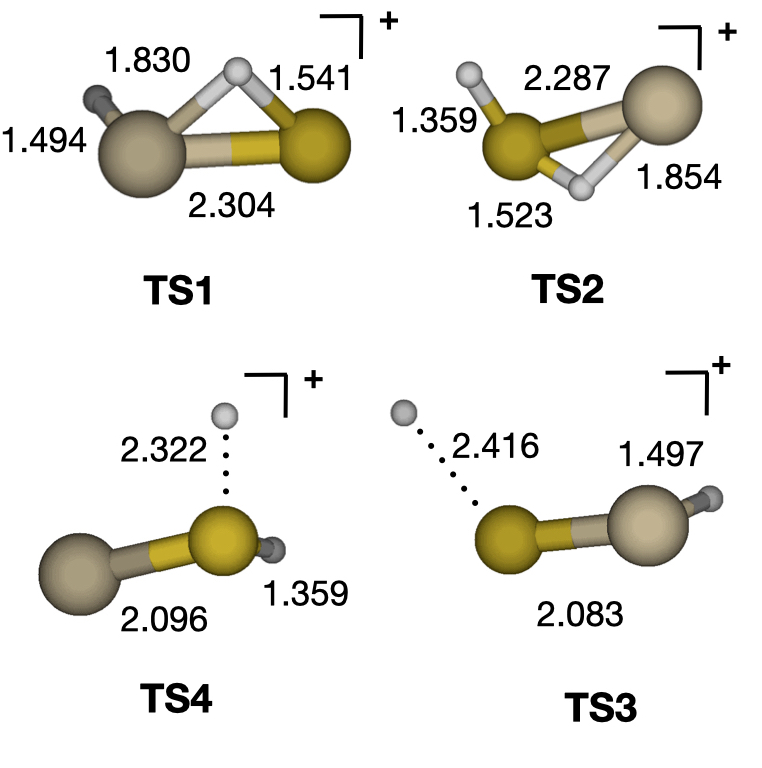}
\caption{Bond distances (reported in Angstroms) for the main transition states of the reaction  SiH$_{2}$($^1$A$_1$) + S$^+$($^4$S), evaluated at the B3LYP/aug-cc-pV(T+d)Z level of theory.  } \label{fig3}
\end{figure}

\begin{figure}[h!]
\centering
\includegraphics[width=0.55\linewidth]{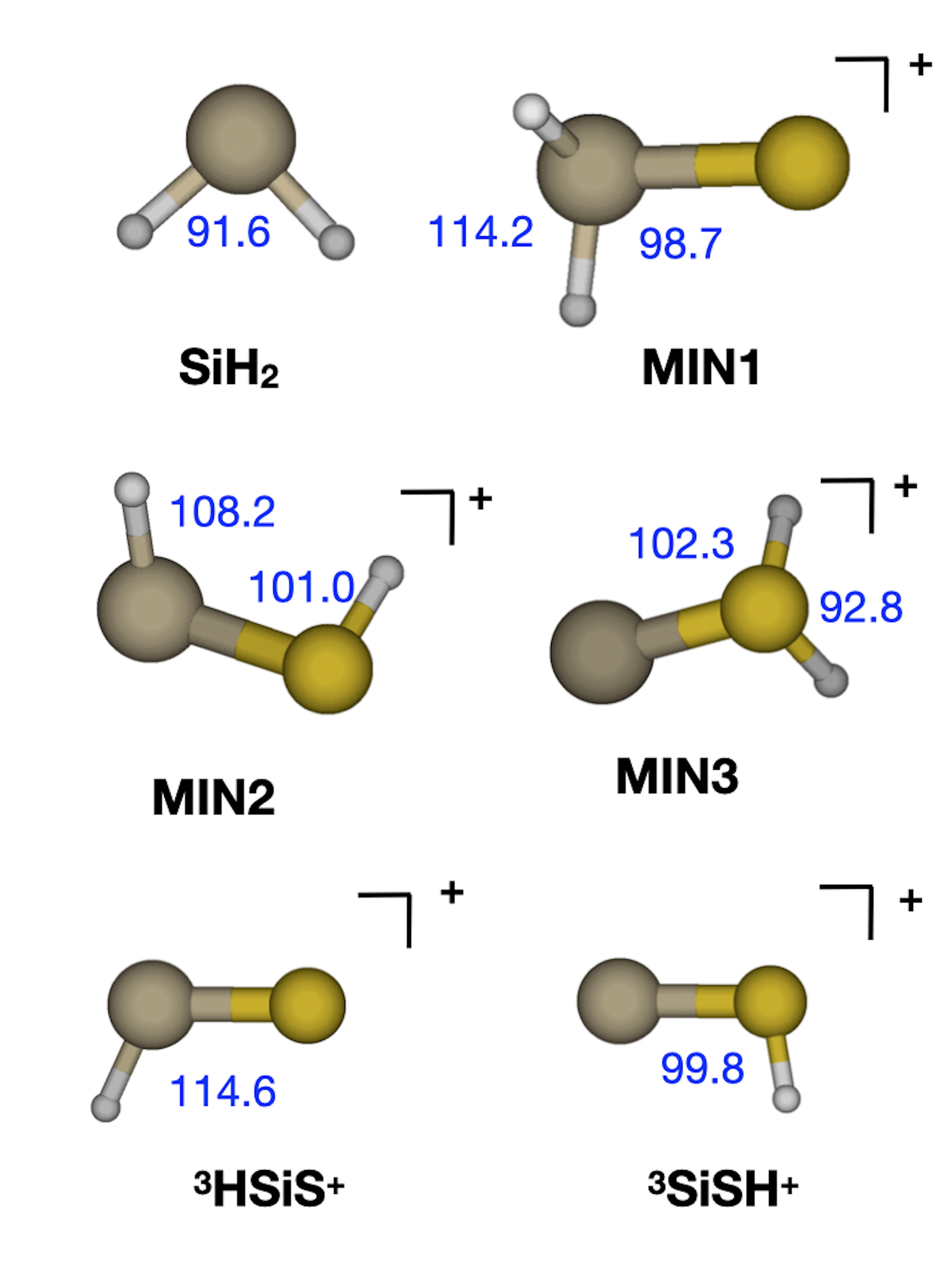}
\caption{Main angles (reported in degree) for the reactant, products and minima identified for the  SiH$_{2}$($^1$A$_1$) + S$^+$($^4$S) reaction at the B3LYP/aug-cc-pV(T+d)Z level of theory.  } \label{fig4}
\end{figure}

\begin{figure}[h!]
\centering
\includegraphics[width=0.55\linewidth]{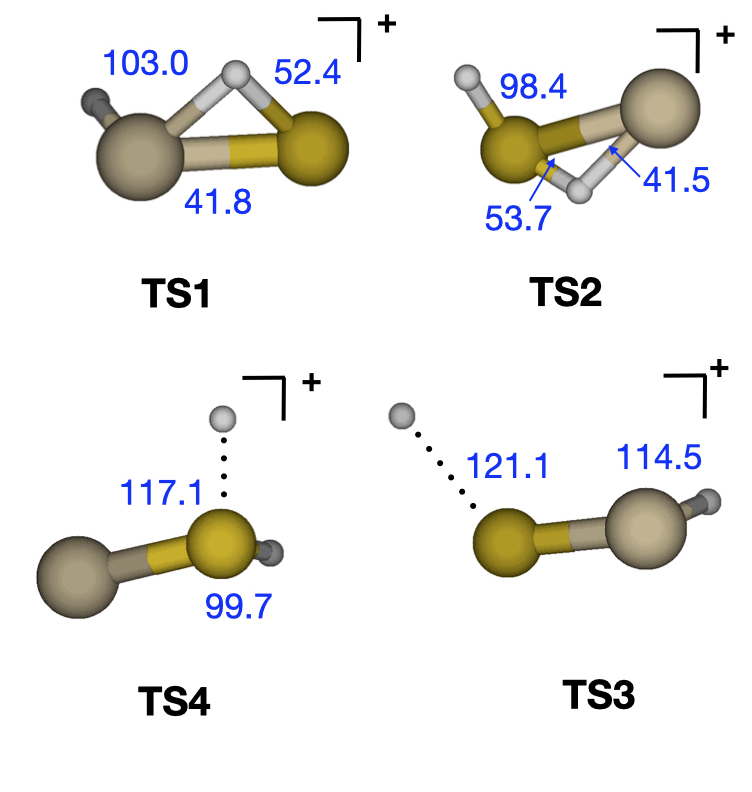}
\caption{Main angles (reported in degree) for the main transition states of the reaction  SiH$_{2}$($^1$A$_1$) + S$^+$($^4$S), evaluated at the B3LYP/aug-cc-pV(T+d)Z level of theory.  } \label{fig5}
\end{figure}

The quartet potential energy surface (PES) for the system SiH$_{2}$($^1$A$_1$) + S$^+$($^4$S) shows three different minima (MIN1, MIN2, MIN3), linked by two transition states (TS1, linking MIN1 and MIN2; TS2 connecting MIN2 and MIN3). The reaction starts with the barrierless interaction of the S$^+$ cation with the SiH$_{2}$ radical, leading to the formation of the intermediate MIN1,located 280.6 kJ mol$^{-1}$ below the reactant energy asymptote. MIN1 features a new Si-S bond with a bond distance of 2.351 \AA. The so formed intermediate can directly dissociate, through a barrierless H-elimination process, leading to the formation of the $^3$HSiS$^+$ cation and an H atom. The global exothermicity of the process is -122.3 kJ mol$^{-1}$. Alternatively, MIN1 can isomerize to the MIN2 intermediate by overcoming a barrier of 93.1 kJ mol$^{-1}$. The related transition state, TS1, clearly shows the migration of a hydrogen atom from Si to S. Two possible combinations of products can be formed starting from the aforementioned MIN2. A barrier of 133.6 kJ mol$^{-1}$ must be overcome in order to form the previously described H+$^3$HSiS$^+$ products. A transition state, TS3, was identified, showing an increase in the S-H bond distance up to 2.416\AA. A different H elimination process can take place starting from MIN2, leading to the barrierless formation of the $^3$SiSH$^+$ cation and an H atom, located 90.1 kJ mol$^{-1}$ below the reactant energy asymptote. The last isomerization process identified in the PES is related to the migration of a hydrogen atom from the Si atom to the S side of the intermediate MIN2, leading to the formation of MIN3, located 204.3 kJ mol$^{-1}$ below the reactant energy asymptote, in which both  H atoms are linked to the S atom of the adduct. A barrier of  99.3 kJ mol$^{-1}$ must be overcome during the migration process. The so formed MIN3 can directly dissociate to H+$^3$SiSH$^+$, overcoming a barrier of 126.7 kJ mol$^{-1}$. The related transition state, TS4, clearly shows the breaking of a S-H bond, with a bond distance of  2.322\AA.
A schematic representation of the PES derived for the SiH$_{2}$($^1$A$_1$) + S$^+$($^4$S) reaction is reported in Figure~\ref{fig1}, while in figures ~\ref{fig2}-~\ref{fig5} the geometry (bond distances in $\AA$ and angles in degree) of the stationary points are illustrated, together with those of reactants and products (at the B3LYP/aug-ccpV(T+d)Z level of theory). A list of enthalpy changes and barrier heights (in kJ mol$^{-1}$, corrected at 0 K) computed at the  CCSD(T)/aug-cc-pV(T+d)Z and B3LYP/aug-cc-pV(T+d)Z levels of theory is reported in Table ~\ref{tab1}.\\
\indent
It is important to stress that, given the multiplicity of the investigated PES, the $^3$HSiS$^{+}$/$^3$SiSH$^{+}$ products are not formed in their ground singlet states, but in the first electronically excited triplet states.

\subsection{The $^3$HSiS$^{+}$/$^3$SiSH$^{+}$ + NH$_{3}$($^{1}$A$_1$) reaction}

\begin{figure}[h!]
\includegraphics[width=1\linewidth]{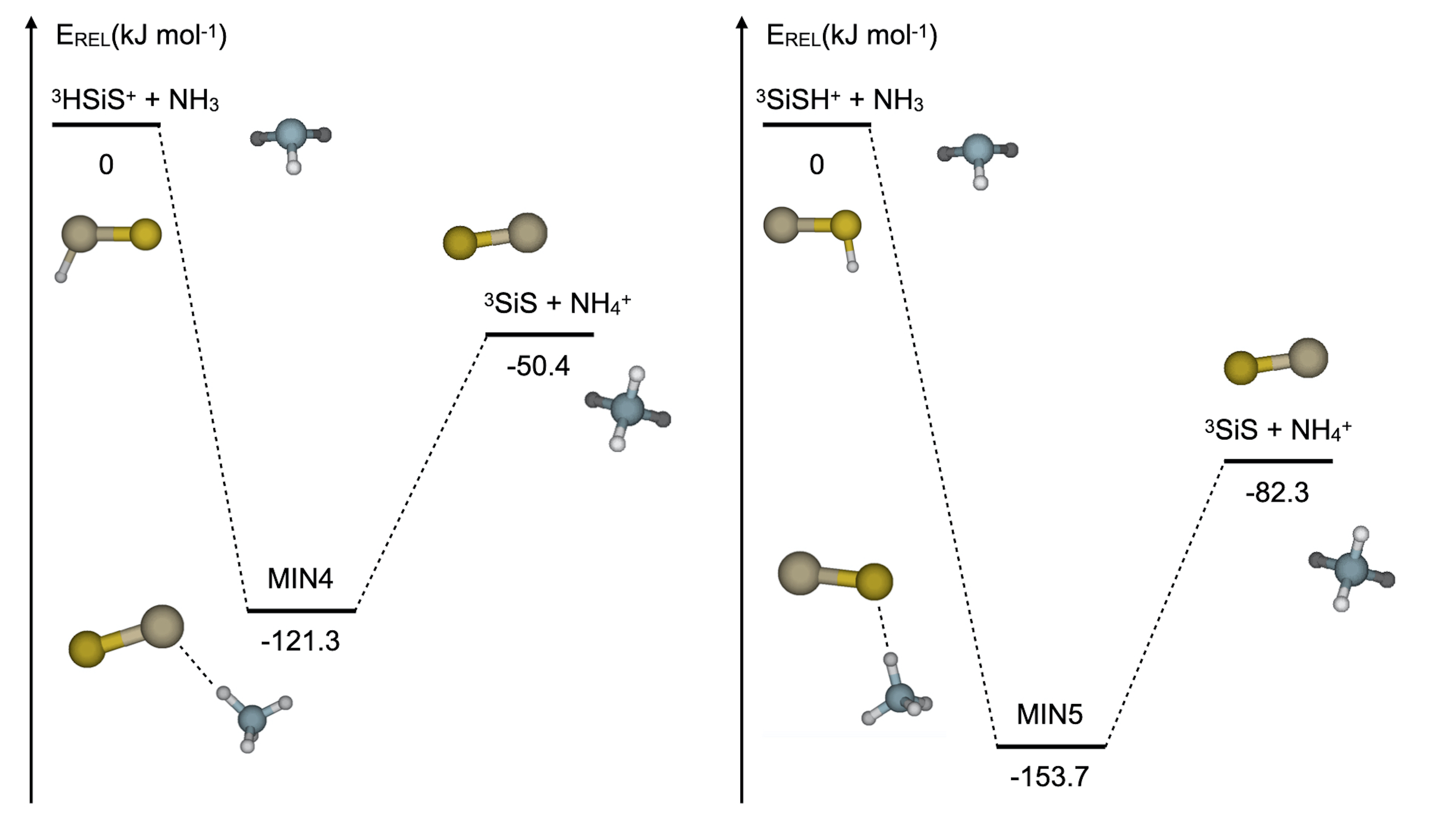}
\caption{2d scheme  of the Potential Energy Surface (PES) for the  proton transfer processes $^{3}$HSiS$^{+}$/$^{3}$SiSH$^{+}$ + NH$_{3}$($^{1}$A$_1$) with the energies evaluated at the CCSD(T)/aug-cc-pV(T+d)Z level of theory. } \label{fig6}
\end{figure}

The barrierless $^3$HSiS$^{+}$ + NH$_{3}$($^{1}$A$_1$) reaction shows a global exothermicity of -50.4 kJ mol$^{-1}$ at the CCSD(T)/aug-cc-pV(T+d)Z level of theory. The interaction between the HSiS$^{+}$ cation and ammonia leads to the formation of the MIN4 intermediate, more stable than the reactants by 121.3 kJ mol$^{-1}$. Once formed, the intermediate can dissociate into  silicon monosulfide ($^3$SiS) and ammonium ion. 
A similar mechanism was derived for the reaction $^3$SiSH$^{+}$ + NH$_{3}$($^{1}$A$_1$). In this case the global exothermicity is 82.3 kJ mol$^{-1}$, while the intermediate MIN5 shows a relative energy of -153.7 kJ mol$^{-1}$ with respect to the reactant energy asymptote. Both the reactions take place without the presence of entrance and/or exit barriers.
The schematic PES of the two processes are reported in Fig. 6, while  the optimized geometries of the most relevant stationary points are reported in Fig.7. A list of enthalpy changes and barrier heights (in kJ mol$^{-1}$, corrected at 0 K) computed at the  CCSD(T)/aug-cc-pV(T+d)Z and B3LYP/aug-cc-pV(T+d)Z levels of theory is reported in Table ~\ref{tab1}

\begin{figure}[h!]
\centering
\includegraphics[width=0.55\linewidth]{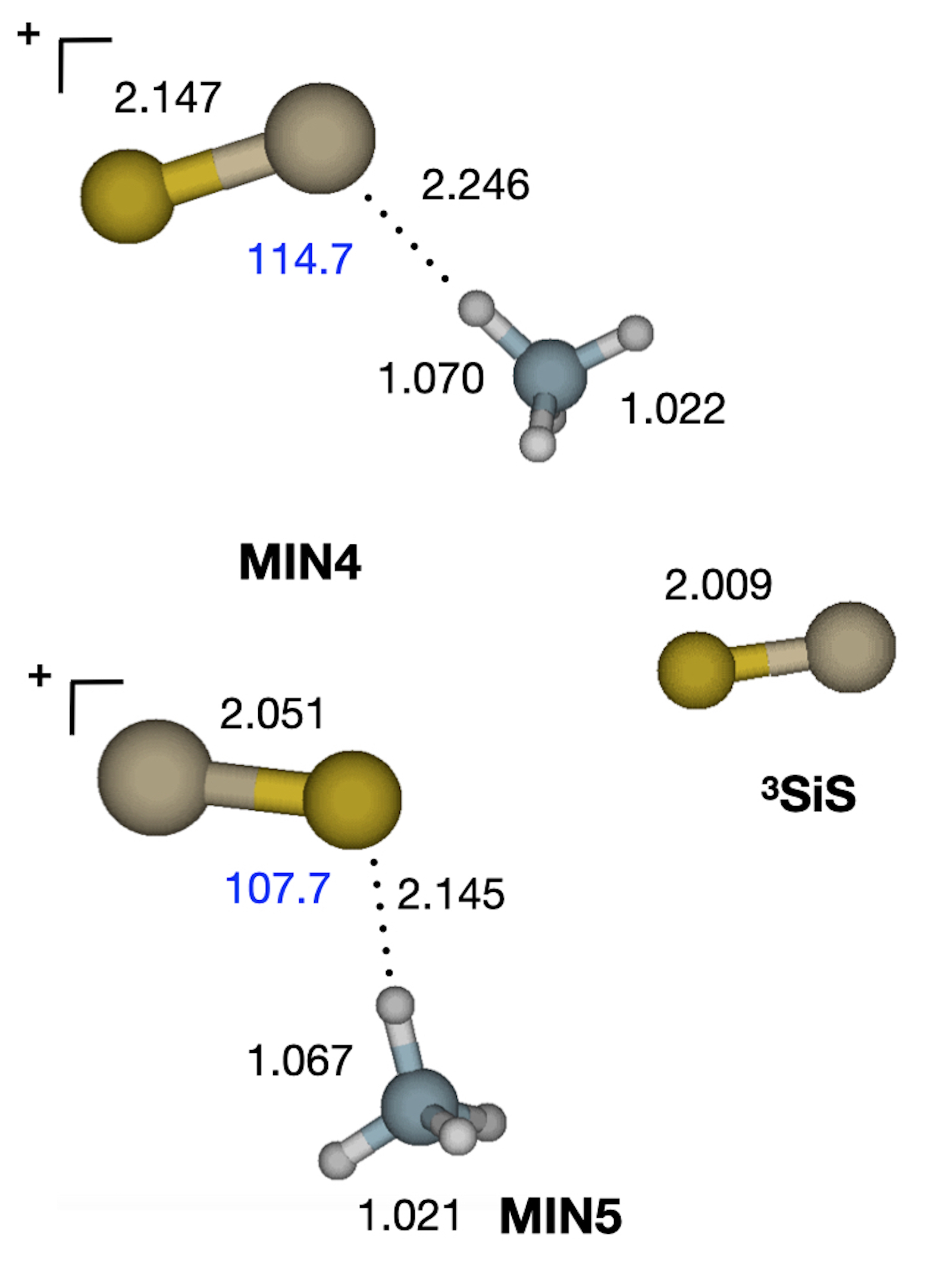}
\caption{Bond distances (in Angstroms) and angles (in degrees) for the main stationary points identified for the  $^3$HSiS$^{+}$/$^3$SiSH$^{+}$ + NH$_{3}$($^{1}$A$_1$) reactions at the B3LYP/aug-cc-pV(T+d)Z level of theory.} \label{fig7}
\end{figure}

\begin{table}[t!]
\centering
\caption{List of Enthalpy variations and computed energy barriers (kJ mol$^{-1}$, 0 K) computed at the  CCSD(T)/aug-cc-pV(T+d)Z and B3LYP/aug-cc-pV(T+d)Z levels of theory for  the system SiH$_{2}$($^1$A$_1$) + S$^+$($^4$S).\label{tab1}}
\begin{tabular}{ |p{4cm}|p{1.5cm}|p{1.5cm}|p{1.5cm}|p{1.5cm}| }
\hline
     & \multicolumn{2}{|c|}{CCSD(T)} & \multicolumn{2}{|c|}{B3LYP}\\ 
\hline
Reaction Path   & $\Delta$H$^{0}_{0}$& Barrier Heights  & $\Delta$H$^{0}_{0}$ & Barrier Heights  \\ 
\hline
SiH$_{2}$ + S$^+$ $\,\to\,$ MIN1    & -280.6  & - & -320.4   & - \\ 
MIN1 $\,\to\,$ MIN2                & 34.5  & 93.1 & 32.2   & 83.9 \\
MIN2 $\,\to\,$ MIN3                & 41.8  & 99.3 & 62.9   & 109.1 \\
MIN1 $\,\to\,$ H + $^3$HSiS$^+$    & 158.3  & - & 150.0   & - \\ 
MIN2 $\,\to\,$ H + $^3$HSiS$^+$    & 123.8  & 133.6 & 117.8   & 120.7 \\
MIN2 $\,\to\,$ H + $^3$SiSH$^+$    & 156.0  & - & 157.8   & - \\
MIN3 $\,\to\,$H + $^3$SiSH$^+$     & 114.2  & 126.7 & 94.9   & 101.2 \\ 
$^3$HSiS$^{+}$ + NH$_{3}$ $\,\to\,$MIN4     & -121.3  & - & -116.4   & - \\ 
MIN4 $\,\to\,$ $^3$SiS +NH$_4^+$     & 70.9  & - & 80.9   & - \\
$^3$SiSH$^{+}$ + NH$_{3}$ $\,\to\,$MIN5     & -153.7  & - & -147.1   & - \\ 
MIN5 $\,\to\,$ $^3$SiS +NH$_4^+$     & 71.0 & - & 71.7   & - \\
 \hline
\end{tabular}
\end{table}

\clearpage

\section{Discussion and Conclusions}

In this work, we have reported the results of a theoretical investigation of the reaction between the S$^+$ cation and the SiH$_{2}$($^1$A$_1$) radical. The reaction has two exothermic channels leading to the isomeric species $^3$HSiS$^{+}$ and $^3$SiSH$^{+}$ formed in conjunction with H atoms. The reaction is not characterized by an entrance barrier and, therefore, it is expected to be fast also under the very low temperature conditions of interstellar clouds. The two ions are formed in their first electronically excited state because of the spin multiplicity of the overall PES.
A subsequent proton transfer process with ammonia can well account for the formation of neutral triplet silicon sulfide, SiS, without invoking electron-ion recombination. \\
\indent
As already noted above, the PES derived for the reaction between the SiH$_{2}$($^1$A$_1$) radical and the S$^+$ cation (in its ground state $^4$S) can only lead to the formation of the two aforementioned radicals in the triplet state, $^3$HSiS$^{+}$/$^3$SiSH$^{+}$. The successive proton transfer carried out by an ammonia molecule (whose ground electronic state is $^1$A$_1$) leads to a triplet potential energy surface. As a consequence, according to the 'correlation rules', 
the title reaction can only form silicon sulfide in a triplet spin state, $^3$SiS. 
However, in systems of this kind, intersystem crossing from the triplet to the singlet PES is certainly possible, being also facilitated by the presence of atoms belonging to elements of the third period of the Periodic Table. In particular, once the first intermediate (MIN1) in the $^3$PES for the S$^+$+ SiH$_{2}$ reaction is formed, an intersystem crossing mechanism can allow reaching the singlet PES. Starting from this point both the $^1$HSiS$^{+}$/$^1$SiSH$^{+}$ cations can be formed and, consequently, the neutral $^1$SiS molecule could be produced.
To account for such an effect, it is necessary to derive the singlet PES (these calculations are currently under way) and to characterize the seam of crossing. Future work from our group will assess the role of intersystem crossing for this system, as already done for a series of reactions involving atomic oxygen \cite{leonori2015experimental,gimondi2016reaction,vanuzzo2016reaction,caracciolo2019combined,cavallotti2020theoretical,vanuzzo2021crossed,fu2012experimental,balucani2015crossed,leonori2012crossed,casavecchia2015reaction}.

As already mentioned in the 'Introduction' section, in the last years different theoretical investigation have been performed to access the viability of gas phase reactions for the formation of interstellar SiS, including a systematic study undertaken by some of the present authors to characterize the possible
formation routes of SiS in low-mass star forming regions via neutral-neutral gas-phase processes\cite{rosi2018possible,rosi2019electronic,skouteris2018theoretical}.
With this contribution we plan to add useful information  to the complex network of gas phase reactions involving Si-bearing molecules. The sputtering of dust grains in shocked regions leads to the release of silicon in the gas phase. Therefore, a final proof of the  unique origin of SiS via neutral-neutral  and/or ion-neutral gas-phase reactions, will allow to use its presence and distribution in interstellar objects as a signpost for that type of chemistry being dominant in the above mentioned regions.

\section{Acknowledgements}

This project has received funding from the Italian MUR (PRIN 2020, “Astrochemistry beyond the second period
elements”, Prot. 2020AFB3FX) and from
the European Union’s Horizon 2020 research and innovation programme under the Marie Skłodowska-Curie grant
agreement No 811312 for the project ‘Astro-Chemical Origins’
(ACO).
    The authors thank the Herla Project - Universit\`{a} degli Studi di Perugia ($http://www.hpc.unipg.it/hosting/vherla/vherla.html$) for allocated computing time. 
    The authors thank the Dipartimento di Ingegneria Civile ed Ambientale of the University of Perugia for allocated computing time within the project “Dipartimenti di Eccellenza 2018-2022”.
   
\clearpage

%
%
%
%

 \bibliography{silicon}{}
 \bibliographystyle{splncs04}

\end{document}